# Impact of substrate temperature on magnetic properties of plasma-assisted molecular beam epitaxy grown (Ga,Mn)N

Katarzyna Gas[a,b,*], Jaroslaw Z. Domagala[b], Rafal Jakiela[b], Gerd Kunert[c], Piotr Dluzewski[b], Edyta Piskorska-Hommel[c,d], Wojciech Paszkowicz[b], Dariusz Sztenkiel[b], Maciej J. Winiarski[d], Dorota Kowalska[d], Rafal Szukiewicz[a,c], Tomasz Baraniecki[c], Andrzej Miszczuk[c], Detlef Hommel[a,c,d], and Maciej Sawicki[b]

[a] Institute of Experimental Physics, University of Wrocław, Pl. Maxa Borna 9, 50-204 Wrocław, Poland
[b] Institute of Physics, Polish Academy of Sciences, Aleja Lotnikow 32/46, PL-02668 Warsaw, Poland
[c] Wrocław Research Center EIT+ Sp. z o.o., Stabłowicka 147, 54-066 Wrocław, Poland
[d] Institute of Low Temperature and Structure Research, Polish Academy of Sciences, Okólna 2, 50-422 Wroclaw, Poland

Keywords: magnetic films and multilayers, nitride materials, crystal growth, magnetic measurements, X-ray diffraction, thin films

**Abstract**

A range of high quality $Ga_{1-x}Mn_xN$ layers have been grown by molecular beam epitaxy with manganese concentration $0.2 \leq x \leq 10\%$, having the $x$ value tuned by changing the growth temperature ($T_g$) between 700 and 590 °C, respectively. We present a systematic structural and microstructure characterization by atomic force microscopy, secondary ion mass spectrometry, transmission electron microscopy, powder-like and high resolution X-ray diffraction, which do not reveal any crystallographic phase separation, clusters or nanocrystals, even at the lowest $T_g$. Our synchrotron based X-ray absorption near-edge spectroscopy supported by density functional theory modelling and superconducting quantum interference device magnetometry results point to the predominantly +3 configuration of Mn in GaN and thus the ferromagnetic phase has been observed in layers with $x > 5\%$ at $3 < T < 10$ K. The main detrimental effect of $T_g$ reduced to 590 °C is formation of flat hillocks, which increase the surface root-mean-square roughness, but only to mere 3.3 nm. Fine substrates' surface temperature mapping has shown that the magnitudes of both $x$ and Curie temperature ($T_C$) correlate with local $T_g$. It has been found that a typical 10 °C variation of $T_g$ across 1 inch substrate can lead to 40% dispersion of $T_C$. The established here strong sensitivity of $T_C$ on $T_g$ turns magnetic measurements into a very efficient tool providing additional information on local $T_g$, an indispensable piece of information for growth mastering of ternary compounds in which metal species differ in almost every aspect of their growth related parameters determining the kinetics of the growth. We also show that the precise determination of $T_C$ by two different methods, each sensitive to different moments of $T_C$ distribution, may serve as a tool for quantification of spin homogeneity within the material.

## 1. Introduction

Dilute ferromagnetic semiconductors (DFS), particularly those in which ferromagnetic (FM) coupling has been realized in single phase material, remain in the focus of research interest since they combine functionalities of semiconductors and magnetic materials, providing a spawning playground for both basic research and technologically viable applications [1]. GaN doped with Mn, (Ga,Mn)N or $Ga_{1-x}Mn_xN$, when $x$ denotes this fraction of Ga atoms that is substituted by Mn, is an important follower of (Ga,Mn)As in the DFS family. From one side a ferromagnetic guise of GaN would constitute a major technological advance due to the already dominating role of group III nitrides in photonics [2] and high power electronics [3]. From the other, the superexchange driven ferromagnetism [4-7] is realized in an insulating host [8,9] opening new, complementary to (Ga,Mn)As [1], avenues for device architectures in which dissipation-less information transfer could be realized [10] or in a combination with piezoelectricity of this wurtzite structure material a direct manipulation of the single-ion magnetic anisotropy specific to $Mn^{3+}$ ions through electric-field-generated uniform strain has been shown possible [11]. Therefore it remains timely and important to work on optimization of the growth conditions aiming at a further increase of Mn incorporation and the magnitude of the Curie temperature ($T_C$), while maintaining the single phase and the insulating character.

From a historical perspective it is fair to say that theoretical predictions based on both the mean-field Zener model [12] and first principle calculations [13] predicting above room temperature FM in modestly Mn doped p-GaN had spurred a considerable worldwide fabrication efforts including resublimation [14], ammonothermal [15], ion implantation [16,17], metal organic vapour phase epitaxy [18], ion-assisted deposition [19], and molecular beam epitaxy (MBE) methods [4,5,20,21]. Encouragingly, contrary to Fe doping, for which there exists a clear solubility limit in GaN of about 0.4% [22], the limit for Mn substitutional incorporation is substantially higher, as single phase samples had been obtained up to $x = 36\%$ [19]. This worldwide technological effort brought a wide spectrum of contradicting magnetic characters ranging from paramagnetic [15,19], through low temperature spin glass [23] and ferromagnetism [4], to ferromagnetic properties seen at room temperature and above [24-26]. The latter is at present widely recognized [27] to originate from regions of spinodal nanodecomposition in the form of either coherent chemical separation or other crystallographic phase precipitates [24,28-30] Since the likelihood of obtaining such a nanodecomposed material depends sensitively on fabrication conditions, it is profoundly important to elaborate dedicated growth protocols either enhancing or eliminating the presence of these high Mn-concentrated volumes in (Ga,Mn)N.

Concerning the MBE growth many different parameters have a profound influence on the resulting characteristics of the layer. The typical constrains with (Ga,Mn)N growth are in many aspects similar to those met with growth of (In,Ga)N compound [31], which in comparison to GaN has to be grown either with a very high nitrogen flux or at sizably reduced growth temperature ($T_g$). However, both approaches have a detrimental influence on the effectiveness of the growth. Assuring high N-flux increases the pressure in the growth chamber what may lead to a drastic reduction of the mean free path down to the limit for the ballistic motion of atoms in the growth chamber. On the other hand, at reduced $T_g$, where the required N overpressure can be reduced, the increased barriers for adatoms diffusion on the crystal surface promote fast island nucleation on atomic terraces leading to a roughening of the crystal growth front. Therefore, since the beginning of the interest in (Ga,Mn)N, the MBE growth effort has been concentrated on probing the Ga/N flux ratio and $T_g$.

For example, the influence of Ga/N flux ratio at a low $T_g$ of 550 °C was investigated by Haider *et al*. [32] It was found that Mn incorporation occurred only either for slightly-metal rich (without Ga bilayer on the surface) or for N-rich growth conditions and the slightly metal rich regime was the optimal one as it resulted in a better surface morphology. Under such conditions up to 5% of Mn was incorporated into a single-phase (Ga,Mn)N. The magnetic properties of those layers were investigated only at room temperature and a clear FM signatures, attributed to metallic precipitates, were detected in samples grown in typical for GaN epitaxy Ga-rich conditions. The similarly strong influence of Ga/N flux ratio on the Mn incorporation was also reported by Kuroda *et al*. [30]. In this study this ratio was varied by two orders of magnitude between the Ga-rich and the N-rich conditions and the most effective Mn incorporation occurred under the N-rich conditions. According to X-ray diffraction up to 2% of Mn could have been incorporated in a single-phase (Ga,Mn)N (at $T_g$ = 720 °C), whereas at higher intended Mn concentrations a presence of secondary phase precipitates was established. Interestingly, the magnetic measurements of one of the dilute (Ga,Mn)N layers ($x$ = 0.3%) revealed FM signatures persisting up to 400 K [33]. In another study by Kocan *et al.* [29] it was found that the growth of homogeneous (Ga,Mn)N layers with $x$ up to 5% could have only been possible at a reduced substrate temperature of 650 °C and under N-rich growth conditions. Above this critical Mn supply second phases precipitates had formed. In the single phase sample the authors did not detect convincing FM signatures what got assigned to a mediocre crystal quality of their layers, which had been grown on Si(111) instead of on high quality GaN templates or pseudo substrates. Instead, the observed weak hystereses at temperatures below 10 K were attributed to a dynamical slow down, ascribed to a low temperature spin-glass phase. On the other hand, it was recently shown [5] that it is possible to grow single phase (Ga,Mn)N layers with Mn concentration as high as 10% at relatively high substrate temperature of 760 °C (optimal for GaN growth in MBE) in metal-rich growth conditions. These layers showed FM with $T_C$ up to 13 K [7].

It has to be underlined, that with the exception of the last three reports [5,7,29], the growth effort of highly concentrated layers, that is with at least a few % of Mn, were not accompanied by a careful and thorough magnetic characterization. In order to shed more light on the issue of a relationship between the growth conditions and the magnetism of (Ga,Mn)N in this report we summarize the results of studies aiming to quantify the role of the growth temperature on the magnetic properties of this material. Additionally, the study has been directed towards the low or very low growth rates, not exceeding 50 nm per hour. It has to be noted that contrary to mastering of the growth conditions of close related nitrides, say (In,Ga)N or (Al,Ga)N, we cannot employ such an important characterization tool as photoluminescence [34], since the mid-gap location of the $Mn^{3+}$ acceptor and the associated Fermi level pining in the mid-gap region [35] serves as the very efficient center for a non-radiative recombination. Therefore our effort is focused on in-depth magnetic studies. We concentrate on the precise determination of the magnitudes of saturation magnetization and $T_C$ and how their magnitudes relate to $T_g$. These studies are supported by an extensive nanoscale structural characterization aiming to probe the possible presence of secondary phases, precipitates or nanoclusters, as well as the chemical phase separation. We employ atomic force microscopy (AFM), secondary ion mass spectrometry (SIMS), transmission electron microscopy (TEM), powder-like X-ray diffraction (XRD), high resolution XRD (HRXRD), as well as synchrotron based X-ray absorption near-edge spectroscopy (XANES) in conjunction with density functional theory (DFT) modelling. The presented results of this orchestrated characterization effort allows us to conclude that solely on the ground of varying only $T_g$ from just below a typical for GaN growth $T_g = 700\,^{\circ}\text{C}$ down to $590\,^{\circ}\text{C}$, *single phase* (Ga,Mn)N layers can be grown at these low growth rates with $x$ spanning a range between 0.2 to 10%. We show that not only does the intended $T_g$ play one of the most decisive roles influencing the magnetic characteristics of the obtained layers, we also prove that the observed variation of the magnetic properties correlate precisely with the actual (local) magnitudes of $T_g$ across the substrate. The latter is attributed to an uneven heat transfer from the external heater to the whole body of the substrate. These previously largely disregarded by the community small variations of temperature over the substrate surface leave a sizable imprint on the magnetic properties which change considerably even over a mere millimeters distances. This finding tells us that regardless the fact that the actual degree of Mn incorporation depends also on other growth conditions, for any reasonable set of parameters the growth temperature alone seems not only to be the most decisive parameter, but that it also allows for extremely fine tuning of $x$ and so other $x$-related properties. Importantly, this remark applies also to other elements required for a growth of ternary or more complex compounds. For example, it has been known for a long time that wafer surface temperature and its uniformity are critical parameters for III-V epitaxy, especially for (In,Ga)N as the In incorporation is highly temperature sensitive. Therefore a suitable means leading to a reduction of $T_g$ fluctuations should be either elaborated or at least their magnitude should be sufficiently accounted for.

    The present studies are intended to provide an additional insight into growth related issues of other multi-ternary compounds, not only from the nitride family, showing in particular a necessity of performing a multi-tool characterization on the same specimens, or, if that is not possible, on as closely located specimens as it can be assured. Particularly, if there exists any indication that the growth temperature may vary across the substrate.

    The paper is organized as follows. After describing the growth process of the investigated layers the results of the structural characterization by AFM, SIMS, TEM, and XRD follow. These measurements prove uniform distribution of Mn ions in the Ga sublattice of GaN. We than investigate the charge state of Mn by XANES with DFT modeling and bring to light $T_g$ distribution on the substrate. After this thorough characterization we present experimental magnetic investigation of the layers obtained from SQUID measurements and in the following discussion section we unravel the connections between $T_g$ and both the Mn incorporation and Curie temperature.

## 2. Experimental

**2.1 Sample Growth**. A Scienta-Omicron Pro-100 MBE chamber equipped with a radio-frequency nitrogen plasma source is used to fabricate all the (Ga,Mn)N layers in this study. The samples are grown on commercially available 3 µm thick GaN(0001) template layers deposited on *c*-oriented 2" sapphire substrates by metal-organic vapor-phase epitaxy. The (Ga,Mn)N deposition is carried out under nominally nitrogen-rich conditions which are very close to stoichiometric when additional Mn flux is provided. The beam equivalent pressures (BEP) of Mn and Ga are set to $0.5 \times 10^{-7}$ and $2 \times 10^{-7}$ mbar, respectively. A nitrogen flux of 1.3 standard cubic centimeters per minute at a plasma power of 400 W is used to provide reactive nitrogen. Such conditions enhance adatom diffusion length allowing to reduce $T_g$ and so the active nitrogen overpressure [36]. It may be understood [37] that this enhancement is assured by a stable Mn overlayer formed at the growth front which additionally provides fast diffusion channels for nitrogen below and for gallium above this Mn layer.

At these growth conditions the deposition rate is quite low, in the range of 20 – 50 nm/h, depending on the growth temperature. The growth time of all the layers is set to 2 hours yielding the (Ga,Mn)N layer thickness (*d*) ranging from 40 to 100 nm, as presented later. Prior to the growth the 2" wafer is diced into four quarters which are individually fixed into a molybdenum frame holder which is heated by an electric heater from the backside. The substrate temperature is governed by the radiative heat exchange because of a lack of a mechanical contact on the backside of the substrate. Its effectiveness is typically assured by metallizing of the back side of the substrate. In our study, following the recipe elaborated previously by some of the present authors a 2 µm thick layer of Ti is used [5]. During the growth, the surface is monitored *in-situ* by reflection high energy electron diffraction (RHEED) and the streaky 1 × 1 pattern (along [11-20]) is observed indicating a smooth two-dimensional (2D) growth (not shown). A series of five samples has been grown at intended growth temperatures ranging from 590 to 700 °C. These temperatures have been set according to the read-out of a video pyrometer with an adjustable focus aiming at the center of the substrate. With a spot diameter of about 3 mm the pyrometer enables to establish the temperature with relative accuracy of about 4 °C. We label the samples as S590, S605, S620, S660, and S700, where the numerical characters indicate the growth temperatures in °C measured at the *center* of the substrate. The issue of the exact temperature distribution over the substrate is addressed separately later in the text.

**2.2 Structural Characterization.** The surface morphology of the layers is investigated by AFM technique using Bruker FastScan set-up at ambient conditions with the samples being mounted directly on a sample stage. The AFM images are acquired with a scan rate of 2.97 Hz at 512x512 data points per image using tapping mode with Bruker TESPA cantilever. The exemplary scans are collected in Fig. 1, and they clearly indicate that the surfaces are homogenous without any droplets or clusters, however an increasing complexity of the surfaces with lowering of $T_g$ is observed. Nevertheless, the hillocks seen on the AFM images exhibit a strongly oblate nature with the height to diameter ratio changing from a mere $4/1500 \simeq 0.0025$ (nm/nm) for sample S700 to about $10/1000 \simeq 0.01$ (nm/nm) for sample S605. Importantly, monoatomic terraces are clearly resolved indicating an overall very low roughness of the growth surface. The corresponding root-mean-square (rms) surface roughness of these layers stays below 2 nm (the exact values are given in the Fig. 1 and in the Table I). On the other hand a somewhat complex surface morphology is observed in S590 sample (Fig. 1e). Here large groups of much narrower and taller hillocks are seen. We connect this evidently larger surface roughness with a sizably reduced surface migration of atoms during the whole growth process at such a low $T_g$ – a conclusion evidenced by the increased density of growth spirals with respect to the other layers. On the other hand, despite their quite spiky appearance, in reality the hillocks still retain an oblate shape, that is the height to diameter ratio remains below $15/170 \simeq 0.1$ (nm/nm). The average rms value for this layer stays below 3.5 nm. Therefore, on the account of the successful incorporation of (Ga,Mn)N layers characterized by rms approaching 8 nm into electrically gated structures withstanding electrical field up to 5 MV/cm [11], we can point out that even at such greatly (as for nitrides) reduced $T_g$ a processing viable material can be obtained. The sufficiently high flatness of this layer for sub-micrometer level of functionalization is further evidenced by TEM presented further below.

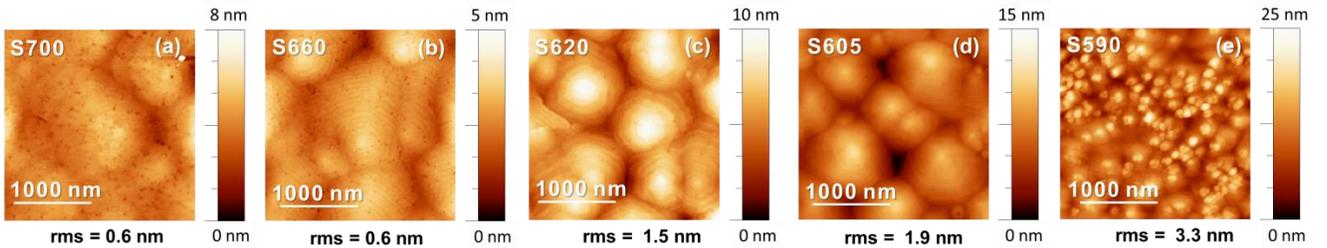

**Fig. 1.** (Color online) $2.5 \times 2.5$ μm$^2$ atomic force microscopy images of the central part of each sample from this study. For each image the corresponding magnitude of the root mean square (rms) is indicated below the frame.

The Mn profiles are established by SIMS using a CAMECA IMS6F microanalyzer. SIMS measurements are performed with oxygen ($^{16}O_4^+$) primary beam at the energy of 8 keV with the current kept at 2 nA. The size of the raster is about $150 \times 150$ μm$^2$ and the secondary ions are collected from a central region of 60 μm in diameter. Mn concentrations are derived from the intensity of Mn+ species, and the matrix signal of N$^+$ is taken as a reference. An Mn-implanted GaN serves as a calibration standard. Mn profiles for all the layers are shown in Fig. 2. There are three main features of these profiles. Firstly, the Mn content changes considerably, particularly for the high growth temperatures. The ratio of $x$ between layers S590 and S700, taken for example at the peak values, is as high as 14. Secondly, the layers' thickness $d$ varies sizably (more than twice between S590 and S700), indicating a greater desorption rate of Ga and Mn species form the growth front at higher temperatures. Finally, two of the layers exhibit a clear depth dependence of $x$, leading to an enhancement of spin inhomogeneity and delectably more smeared Curie transition, as discussed later in the magnetic part of the paper. Magnitudes of $x$ and $d$ established from these profiles are listed in Table I.

Two samples grown at the lowest temperatures, S605 and S590, are subjected for TEM investigations. The S590 one exhibits the roughest surface in the whole range of samples, so it has been particularly likely that this layer may possess the highest amount of structural defects. The "next" layer,

S605, grown at slightly higher $T_g$ is meant to serve as a reference one for the S590. The TEM studies are performed in FEI Titan cube 80-300 image aberration corrected microscope operating at 300 kV. Cross-sectional electron-transparent specimens for TEM investigations are prepared in the form of lamellas protected by platinum layer and mounted on Cu grid applying standard FIB technique.

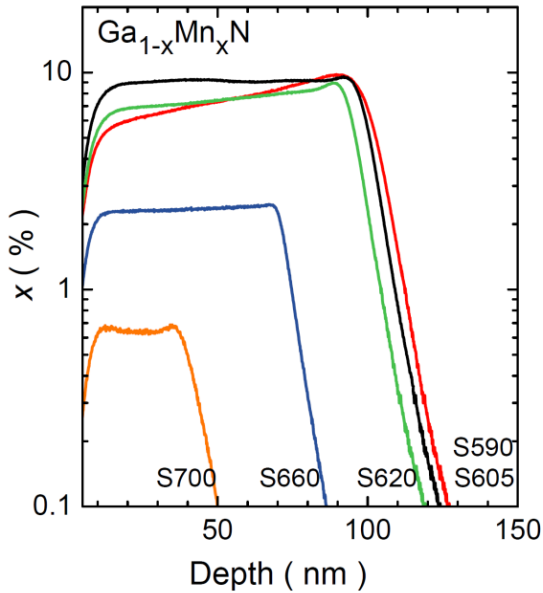

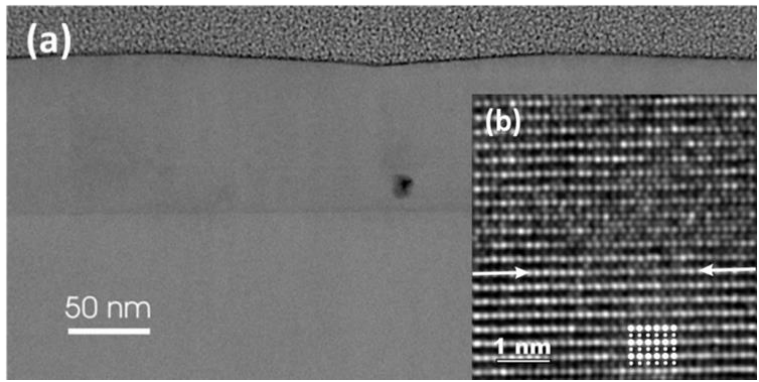

**Fig. 2.** (Color online) Secondary ion mass spectrometry Mn-depth profiles measured on pieces cleft form the central region of each layer.

**Fig. 3. (a)** High-angle annular dark-field/scanning transmission electron microscopy image of layer S590 shows a uniform contrast of the (Ga,Mn)N layer except sporadically observed a few nanometer in size dark spots originated from voids. The top surface undulation reaches an amplitude of about 7 nm. **(b)** A high resolution image of the (Ga,Mn)N/GaN interface of this sample. The horizontal arrows indicate atomically flat interface. The inset shows the $[\bar{1}100]$ projection of GaN crystal lattice (the big and small circles represent Ga and N atomic columns, respectively).

A high-angle annular dark-field/scanning transmission electron microscopy image of layer S590 exemplifying a very high homogeneity of the investigated material is shown in Fig. 3a. In particular, no indications of phase separation or cluster incorporation is found. The (Ga,Mn)N/GaN(buffer) interface is sharp within one atomic layer, as presented in Fig 3b. Mn is incorporated in a dilute manner solely in the wurtzite (Ga,Mn)N phase - Mn atoms substitute Ga atoms with neither noticeable ordering nor segregation. Both images confirm that this layer exhibits a perfect bulk-like appearance. The main departure from the perfect planar growth is seen at the top-most part of the layer, where a weak surface undulation is seen. The observed amplitude of the undulation (crest to trough distance) of about 10 nm compares well to the average value of 15 nm established by AFM. The other deviation from an ideal structure is the presence of rare, nanometer in size voids, seen at the central part of Fig. 3a. Interestingly, as presented later in this study, none of these features exerts any particular influence on the magnetic properties of this layer. A rough quantitative Energy-dispersive X-ray spectroscopy (EDX) analysis yields $x = (8 \pm 1)\%$ for this sample, what compares very well with values established by more precise methods, summarized in Table I. Qualitatively the same structural properties are found for layer S605 with only this exception that the surface warping is weaker, again, in hand with the AFM findings.

The TEM images reveal also the presence of threading dislocations (TDD) piercing the whole (Ga,Mn)N/GaN(buffer) stack, as evidenced in Fig. 4. The detailed EDX profiling reveals a sizably reduced Mn content in the dislocation core area. This low Mn content found along the dislocation can explain why the TDD can remain electrically active, effectively shunting vertical transport across (Ga,Mn)N layers [38,39], whereas the high-$x$ surrounding material retains highly insulating properties [8,11,40].

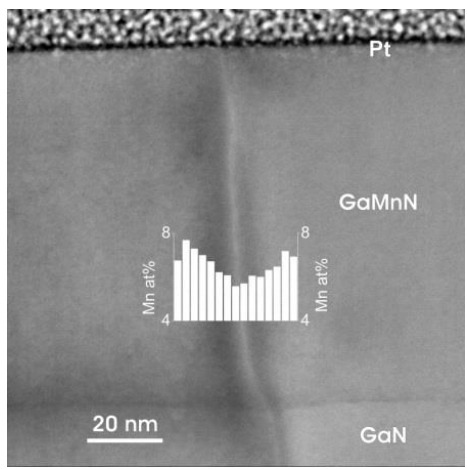

**Fig. 4.** High-angle annular dark-field/scanning transmission electron microscopy image of S590 layer with a treading dislocation. The inset shows a reduction of Mn concentration in a region of the dislocation core.

Crystal structures, epitaxial relationships, in- and out-of-plane lattice parameters, and interface quality are further determined by X-ray diffraction using two laboratory Philips X-ray diffractometers (wavelength of $Cu_{K\alpha1} = 1.5406$ Å). First, for powder-like diffractometry, a Bragg–Brentano X-ray diffractometer (X'Pert MPD) equipped with a Johansson Ge(111) monochromator and linear position sensitive detector (the setting are described in Ref. 41) dedicated to search for other phases of manganese compounds. Second, to characterize structural quality and strain condition of the (Ga,Mn)N layers, a high resolution (HR) X'Per MRD diffractometer with an incident beam formed by an X-ray mirror and monochromator [four-bounce (220) Ge asymmetric], and diffracted beam detected directly by point detector or indirectly through three-bounce Ge(220) analyzer crystal. Samples are mounted on a high precision goniometer stage.

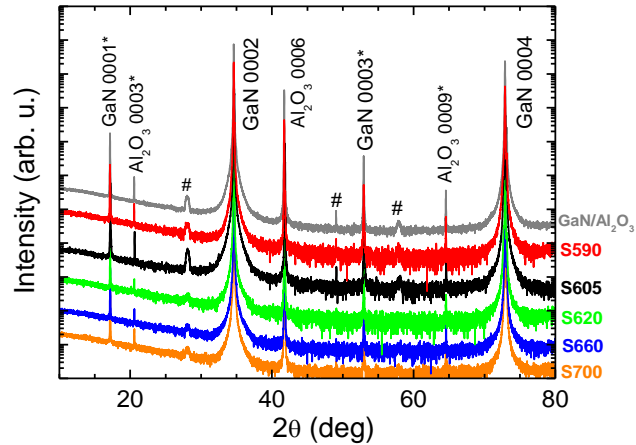

**Fig. 5.** (Color online) X-ray diffraction pattern 2θ of all (Ga,Mn)N layers in this study and of a reference GaN template. The scans are shifted vertically for clarity of the presentation. Within a dynamic range of intensity exceeding four orders of magnitude no reflections specific to other phases inclusions could have been detected. Features marked by # or * are experimental artifacts.

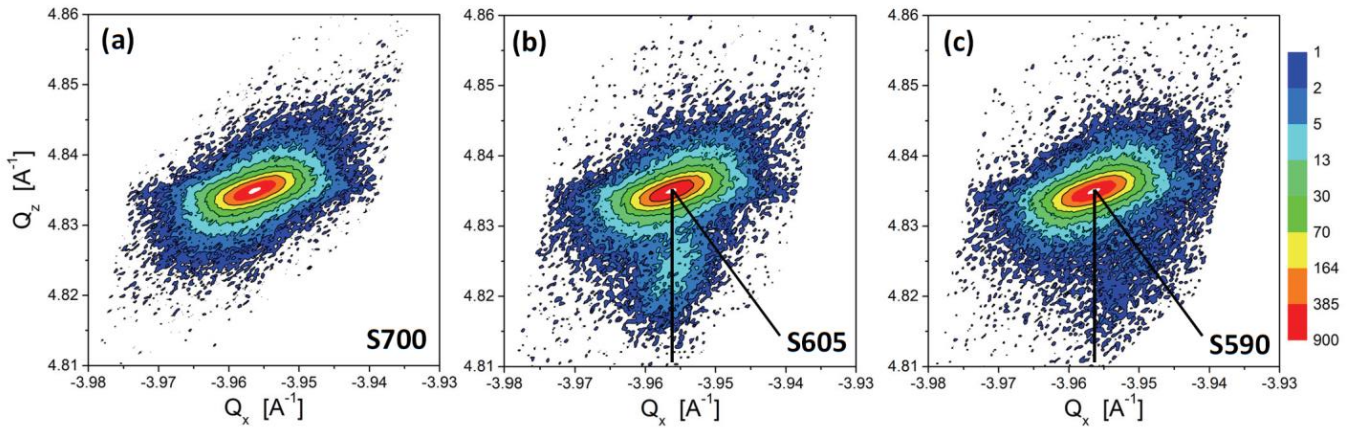

**Fig. 6.** (Color online) Reciprocal space maps of the asymmetric $\bar{1}\bar{1}24$ reflection. The lines vertical and inclined to the $Q_x$ axis [(**b**), (**c**)] show the triangle of relaxation. Depending on the degree of the layer relaxation, the signal from the layer should lie between these lines (fully strained – vertical line goes through it, fully relaxed – the inclined line intersects the node). The high intensity node is related to the GaN template layer. The low intensity node as in (**b**) origins from the fully strained (Ga,Mn)N layer.

The search for alien crystallographic compounds is performed in spinning mode in the wide 2θ range (from 10° to 80°) for all the samples from the study, including the reference templated substrate. The samples are oriented along the growth direction of GaN [0001] covering the symmetric 0002 and 0004 Bragg diffraction peaks of GaN buffer, which are the main features in Fig. 5. The features marked with # or * are experimental artifacts [42], notably they are all seen in the scan obtained for the GaN template test sample. These scans confirm the high crystallinity of the epitaxial layers and within the sensitivity of the method exclude the presence of precipitations of secondary phases such as $Mn_4N$, $Mn_3Ga$, Mn, $Mn_3GaN$, and $Mn_{1-x}Ga_x$ - the most frequently observed precipitations in MBE-grown (Ga,Mn)N layers [27], and so confirming the conclusions of the previously discussed TEM studies.

The degree of a modification of the GaN lattice parameter due to the incorporation of Mn is assessed from symmetric 0002 and asymmetric $\bar{1}\bar{1}24$ (so called "high incidence") reflections collected by the HR diffractometer. The initial assumption of the pseudomorphic growth of so thin (Ga,Mn)N layers (several tens of nanometers) is verified by the reciprocal space map of the $\bar{1}\bar{1}24$ reflection, Fig. 6, where the three most characteristic cases are exemplified. Two main features can be recognized there: the high intensity peaks located at $Q_z = 4.835$ Å$^{-1}$ – related to the GaN template, and a much lower intensity peak at smaller reciprocal lattice units from the thinner (Ga,Mn)N layer. However, the position of the latter is visible very clearly only for layers S605 and S620 (not shown). The same $Q_x$ of the (Ga,Mn)N and of the GaN buffer peaks indicates perfect pseudomorphic growth of this layer. The intensity of the second peak in layer S590 is smeared towards the line indicating full relaxation, a sign that partial strain relaxation takes place in this layer. We come back to this issue below. The lack of the second peak for layer S700 (Fig. 6a) and S660 (not shown) is related to both too small $x$ and $d$ in these layers.

To quantify the above mentioned features we collect HR 2θ/ω scans for the 0002 Bragg reflections for all investigated samples, Fig. 7. The high intensity peaks at 34.54° are related to the GaN template layer whereas the lower intensity and very broad peaks at smaller angles origin from the much thinner (Ga,Mn)N layers. Their position shifts with decreasing $T_g$ to smaller angles, i.e., to larger lattice parameters indicating increasing Mn content. The existence of X-ray interference fringes, best seen for S620 layer, implies high structural quality of the layers and confirms the established by TEM very good quality of the interfaces. It has to be noted that the lack of clearly resolved "thickness" oscillations for the two layers with the smallest $x$ (S700 and S660) is most likely due to too small difference of the refractive index between GaN and very low-$x$ (Ga,Mn)N and does not necessarily have to indicate their lesser quality. Thin black solid lines in Fig. 7 represent results of simulation performed using commercially available PANalytical EPITAXY software based on the dynamical theory of X-ray diffraction. For the calculations the same Poisson's ratio of 0.202 [43] and Debye-Waller factor 0.7 [44] are assumed for both (Ga,Mn)N and GaN layers. Lattice parameters for GaN ($a = 3.1885$ Å, $c = 5.1850$ Å) are taken from Leszczynski *et al.* [45] and the hypothetical MnN wurtzite structure lattice parameter is after Kunert *et al.* [5] The generally small magnitudes of $x$ and $d$ do not permit to obtain a perfect match between the experimental scans and the simulations. The best fit has been obtained for layer S620, for which the oscillation fringes are best resolved. For the other layers, the simulations are mainly guided by the intensity of the side peak from the layer. For example, the layer thickness can be simulated out by establishing the best account for the asymmetry of the diffraction curve, since the size of "hump" on the *left* to the main GaN peak scales with the width of the layer. The procedure turns out to be effective enough as the layers' thicknesses obtained from the simulations agree within 10% with the values derived from TEM and SIMS measurements. The other two parameters needed to be specified for the simulations of the 0002 reflection curve are the Mn content and the relaxation degree ($R$) of the layer to GaN. $R \equiv 100\%$ for the fully relaxed layer. However, as seen in Fig. 7, the worse match is obtained for layer S590 despite having, according to SIMS, larger $x$ and similar $d$ as S620, for which experimentally resolved thickness fringes can be perfectly reproduced. There are two effects which contribute. The first is a partial layer relaxation. Indeed, the most reasonable fit to the experimental data has been obtained assuming $R = 80\%$. The second is the presence of small angle surface undulations which smear out the Braggs' interference condition and smooth out the diffraction pattern. On the same token, the HRXRD pattern of sample S605 may signalize that it has been grown just on the border condition for optimum (Ga,Mn)N growth at these low growth rates as are exercised in this study. Finally, upon the simulations of the 0002 reflection the magnitude of the wurtzite $c$ lattice parameter is established and using information from reciprocal space map (reflection $\bar{1}\bar{1}24$, as in Fig. 6) the value of the basal plane parameter $a$ is obtained. However, the inability of the accurate assessment of S590 layer's maximum node position (Fig. 6c) denies a sufficiently high accuracy determination of the magnitude of $a$ for that sample. The established magnitudes of $x$ and $d$ are listed in the Fig. 7.

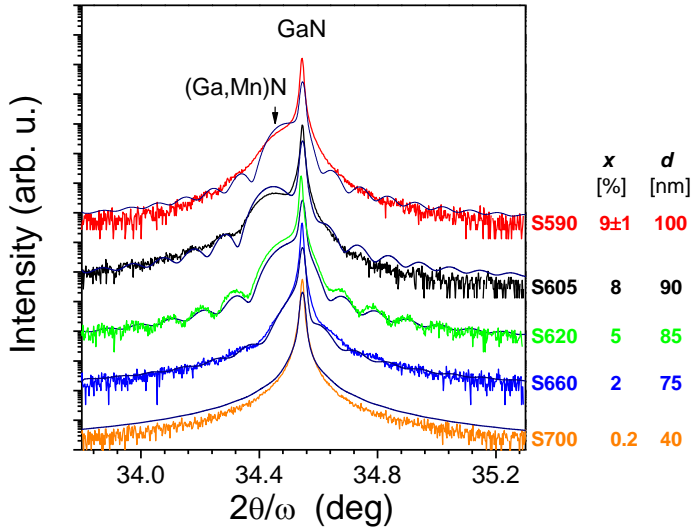

**Fig. 7.** (Color online) Thicker lines of various shadows: high-resolution X-ray diffraction patterns of the studied layers: 0002 Bragg reflections, 2θ/ω scans for the (Ga,Mn)N layers. The central narrow features correspond to reflections from the GaN buffer layer and the broader peaks at lower angles are the reflections from (Ga,Mn)N layers. Mn contents and layer thicknesses quoted aside to the corresponding scan are obtained from simulations, marked in the figure by thin solid black lines.

X-ray Absorption Fine Structure (XAFS) at the Mn $K$ absorption edge (6539 eV) are performed at the SAMBA beamline at the synchrotron SOLEIL in Saint-Aubin, France. The synchrotron radiation beam is monochromatized using a Si(111) monochromator. The Mn $K$ absorption spectra for S620 (Ga,Mn)N layer and the three Mn-oxides references containing Mn ions in +2, +3 and +4 oxidation states have been acquired at room temperature in the fluorescence mode. The obtained spectra after a standard normalization performed within the Athena program [46] are shown in Fig. 8a. There are two main features of the $K$-edge XANES spectrum which are indicative of the Mn valence state. They are (i) the position of the absorption edge, which, in accordance with the Haldane-Anderson rule, is known to move to higher energy with the increasing charge state, and (ii) the presence of a pre-edge feature which singlet or doublet structure discriminates between +2 [47-49] or +3 [18,48-50 Mn oxidation state, respectively. The structure is a consequence of the dipole allowed transitions from Mn $1s$ core levels to $p$-symmetry orbitals. Mn atoms in wurtzite (Ga,Mn)N are tetrahedrally coordinated, so the Mn $3d$ levels split into two nearly degenerate *e* and three nearly degenerate $t_2$ levels for each spin direction. The tetrahedral environment of Mn atoms permits 4*p*-3*d* hybridization of Mn states and thereby a dipole transition of *s* electrons to the *p* components of the hybrid orbitals is allowed. Therefore, the doublet form of the pre-edge structure results from two electronic transitions from the Mn 1*s* localized orbital to the lower-energy $t_2$ partially occupied anti-binding hybrid *p-d* orbital. The low energy peak is the transition to the $t_2$ spin-up orbital, whereas the higher energy one is to the $t_2$ spin-down orbital [47,51].

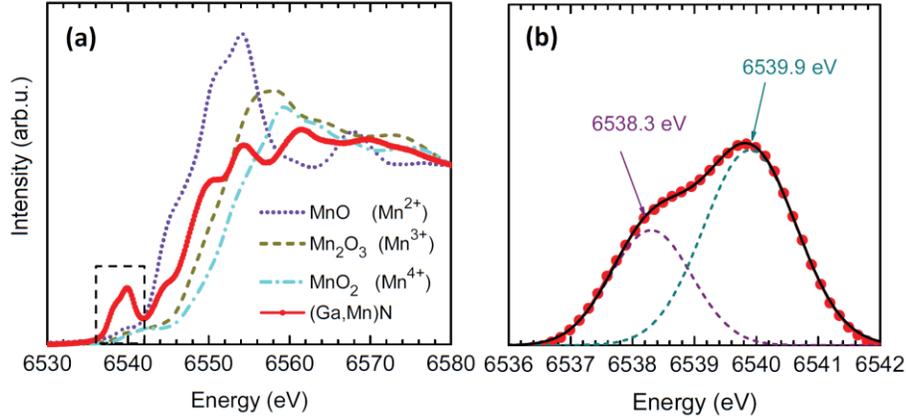

**Fig. 8.** (Color online) **(a)** Normalized X-ray absorption spectra near the Mn *K* absorption edge of the investigated (Ga,Mn)N layer (bullets) and three reference oxides representing Mn in +2, +3, and +4 valence (lines). **(b)** Blown up part, marked by a dashed rectangle in (a), of a pre-edge part of the (Ga,Mn)N layer's spectrum after correcting the experimental spectrum for an edge baseline (bullets). The solid lines represent the fit of the pre-edge structure by a superposition of two Gaussian components indicated by two separate dashed lines.

We start the analysis from focusing on the position of the Mn *K* absorption edge. It can be concluded from Fig. 8 that the absorption edge of the (Ga,Mn)N layer locates itself between MnO ($Mn^{2+}$) and $Mn_2O_3$ ($Mn^{3+}$), perhaps much closer to the latter one if the broad feature at the center of S620 layer absorption edge is disregarded. Therefore, strongly favoring +3 Mn oxidation state. The dominance of the +3 oxidation state of Mn ions in the studied layer is further supported by a clear observation of a complex pre-edge structure. Indeed, as it is indicated in Fig. 8b this structure is composed of two peaks. We model this asymmetric shape by a superposition of two Gaussian components centered at 6538.3 and 6539.9 eV, which separation, ~1.6 eV, fits well the range of 1.5 – 2.0 eV found in other studies [18,46,47,52].

More insight into the Mn oxidation state can be deduced from modeling of our XANES experimental spectra. This is done by the full-potential linearized augmented plane wave (FLAPW) Wien2k package [53] using the Perdew-Wang parametrization of the local density approximation (LSDA) for the exchange-correlation energy [54]. The calculations are performed for a wurtzite-type supercells consisting of 32-atoms, i.e., the 2×2×2 multiplicity of a primitive cell. The experimental values of the lattice parameters, *a* = 3.180 Å, *c* = 5.166 Å, and a free atomic position at *z* = 0.375 are adopted. The parent $Ga_{15}MnN_{16}$ system in which 1 out of 16 Ga atom is substituted by an Mn atom (this corresponds to x ≃ 6.3%, well matching to our sample), and modified structures with a nitrogen vacancy ($Ga_{15}MnN_{15}$) or an oxygen atom ($Ga_{15}MnN_{15}O$) in the nearest neighboring nitrogen site to a Mn atom are considered too. The valence basis cutoff of 8 eV, and the 6 × 6 × 6 *k*-point mesh is employed. The results of the theoretical predictions are compared with the experimental findings in Fig. 9. We find that all features of the experimental spectrum (labeled as *A*, *B*, *C*, *D*) depicted in Fig. 9a are present in the theoretical spectrum for the ideal $Ga_{15}MnN_{16}$ system, shown in Fig. 9b. The obtained by us spectrum corresponds very well to that reported previously by Titov *et al.* [46]. Most importantly, the experimentally established shape of the pre-edge double structure is captured correctly. The analysis of the spin-up and spin-down contributions into the total spectrum suggests that the pre-edge peak is formed by partially empty Mn 3*d* spin-up and completely unoccupied Mn 3*d* spin-down bands.

On the other hand the origin of the experimental feature B is somewhat unclear. On one hand, an overall downshift of the conduction bands, which is characteristic for approaches based on DFT, causes a mixing between the features *A* and *B*, and as a result, the total spectrum is artificially flattened in this region. On the other hand, the feature *B* may be related to some modifications of the electronic structure of (Ga,Mn)N, which are caused by a presence of nitrogen vacancies or oxygen atoms in the vicinity of the Mn atom. As presented in Fig. 9c, the whole spectrum of the model $Ga_{15}MnN_{15}$ system with the nitrogen vacancy is strongly flattened when compared with that of the parent $Ga_{15}MnN_{16}$ material. Furthermore, the deficiency of a nitrogen atom leads to the clearly singular pre-edge structure *A*, which is not experimentally observed. We note also that the presence of an oxygen atom in the $Ga_{15}MnN_{15}O$ material leads to the theoretical spectrum, depicted in Fig. 9d exhibiting a clear singular pre-edge structure as expected for $Mn^{2+}$ ions, but located at lower energy than that for $Ga_{15}MnN_{15}$ system. Therefore we conclude that the findings presented here suggest that the standard DFT calculations describe well the pre-edge structure of the high quality (Ga,Mn)N indicating a mixed valence of +3 and +2 of Mn states, but having decisively predominantly +3 character.

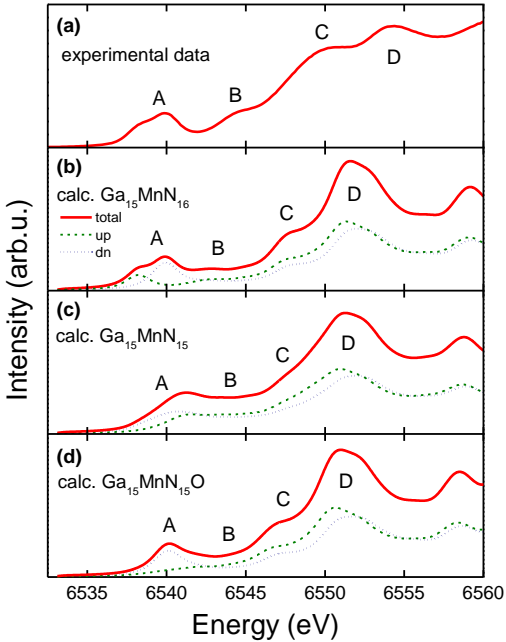

**Fig. 9.** (Color online) Comparison of the experimental X-ray-absorption near-edge structure Mn *K* spectrum of the S620 (Ga,Mn)N layer **(a)** and theoretical predictions obtained using methods based on density functional theory for $Ga_{15}MnN_{16}$ **(b)**, $Ga_{15}MnN_{15}$ **(c)**, $Ga_{15}MnN_{15}O$ **(d)** systems. The total (red solid line), spin-up (green dashed line), and spin-down (blue dashed line) spectra are marked.

Finally, we turn to a more precise determination of the temperature distribution over the substrate surface during the growth. As indicated in Fig. 10 for S620 layer the temperature across the substrate indeed varies considerably. The data indicate that the temperature changes from the edges to the center by more than 10 °C, where actually the lowest $T_g = 620$ °C is recorded. In this particular case the substrate heater was heated to 830 °C. The presence of this radial gradient is generally expected since even a 2 µm thick Ti layer cannot absorb the thermal radiation of the wavelength of about 2.6 µm as effectively as an about 1 mm thick Mo frame holder does, and so a sizable portion of the heat flows to the substrate from the edges which are in mechanical contact with the more effectively heated Mo holder. Therefore, in the whole study the samples (growths) are labeled according to the magnitude of $T_g$ recorded at the central part of the substrate, however, the more detailed discussion relies on $T_g$ values established for different zones on the substrate, corresponding to the particular, small pieces cleft from their parent substrate. Very importantly, whenever proved possible both the magnetic and structural characterization are performed on physically the same small specimens.

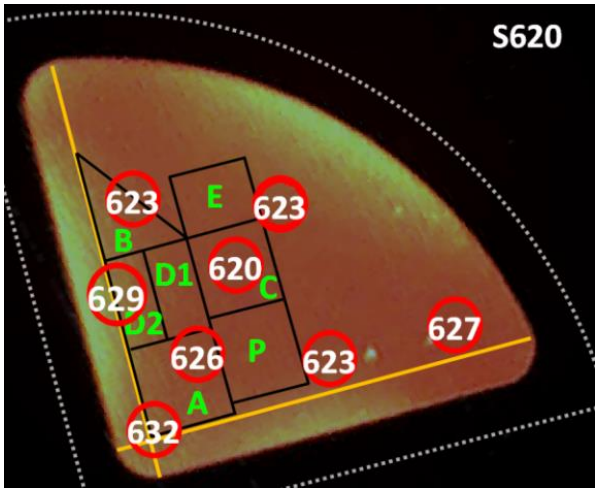

**Fig. 10.** (Color online) Temperature distribution across the substrate established before commencing the growth of the sample S620. About 2 mm wide outer rim of the substrate is hidden behind the holder, as indicated by the dashed line. The approximate positions at which the pyrometer was aiming are indicated by the circles and the corresponding readouts are given in °C. The approximate position of specimens A, B, C, D, E, and P are indicated by the black lines. Please note that to avoid so called edge effects the samples are cut several mm from the border of that part of the substrate on which the layer was deposited (marked by light solid lines)

**Table I**. Data related to the investigated (Ga,Mn)N layers. The following values are listed: the growth temperature ($T_g$), code used for the sample, root-mean-square (rms) surface roughness, thickness ($d$) and Mn content ($x$) established by SIMS, basal ($a$) and vertical ($c$) hexagonal lattice parameters established by XRD, the Mn concentration ($x_M$) as obtained from magnetization data, and two Curie temperatures established by TRM ($T_C^{max}$) and inflection point ($T_C^{mean}$) methods.

| $T_g$ (°C) | Sample name | rms (nm) | SIMS d (nm) | SIMS x (%) | XRD a (Å) | XRD c (Å) | $x_M$ (%) | SQUID $T_C^{max}$ (K) | SQUID $T_C^{mean}$ (K) |
|---|---|---|---|---|---|---|---|---|---|
| 590 | S590 | 3.3 | 95 | 7.3 | 3.1902 | 5.1980 | 6.8 | 13.1 | 9.3 |
| 605 | S605 | 1.9 | 95 | 8.4 | 3.1849 | 5.2031 | 7.8 | 13.6 | 10.5 |
| 620 | S620 | 1.5 | 90 | 7.2 | 3.1847 | 5.2000 | 6.2 | 10.5 | 6.3 |
| 660 | S660 | 0.6 | 70 | 2.2 | 3.1849 | 5.1936 | 2.1 | < 2 | < 2 |
| 700 | S700 | 0.6 | 40 | 0.6 | - | - | 0.3 | < 2 | < 2 |

**2.3 Magnetic Properties Measurements.** For the magnetic measurements samples are cut into approx. (4 × 5) mm$^2$ specimens, as it is depicted in Fig. 10, and are washed in concentrated HCl in order to remove the Ti thermalization back layer and possible other traces of ferrous contaminants deposited on surfaces and edges during handling. The latter is regarded very unlikely, as all the specimen handling is done only using dedicated plastic tools in clean and Fe contamination-free environments. The magnetic measurements are performed in a Quantum Design MPMS 7 T superconducting quantum interference device (SQUID) magnetometer equipped with a low field option following strictly the guidelines of precise magnetometry of thin layers on a substrate [55,56]. We perform the measurements in two regimes of the magnetic field. The high field measurements enable to establish the general shape of the field dependence of magnetization, $M(H)$. At high temperatures $M(H)$ yields about a possible presence of high-$T_C$ ferromagnetic precipitates and other signatures of a significant nonrandom Mn distribution within the host. At low temperatures an independent assessment of $x$ through the saturation of $M$, a *magnetic* concentration ($x_M$), is possible. Measurements performed in weak field regime are performed to assess the $T_C$.

There are two challenges associated with the experimental determination of magnitude of $x_M$ in (Ga,Mn)N. The first one is related with the sufficiently accurate determination of the magnitude of $M(H)$. We note that (Ga,Mn)N layers are typically grown on about 300 - 400 μm thick sapphire (or other suitable material) substrates so that the Mn-doped layers constitute only a tiny fraction of the volume of the whole specimen, and due to the substantial magnetic dilution their magnetic moment is small compared to the diamagnetic signal of the substrate, particularly at strong magnetic fields. As these substrates contain a certain amount of (para-)magnetic dopants [22], their magnetic response at lowest temperatures is distinctly far from an ideal diamagnetic one, in particular being nonlinear with $H$ and, moreover, temperature dependent [22]. Although these nonlinearities can be regarded as weak, in this case their magnitude is at least comparable to the magnitude of the researched signal of the (Ga,Mn)N layers. Therefore, following the guidelines outlined in refs. 55 and 56 the magnetic data presented in this paper are obtained after subtracting the magnetic response of a reference sapphire substrate specimen cut from the same batch of substrates on which the investigated set of samples is grown and independently measured on the same holder and according to the same experimental procedure. The precise value of the scaling factor for the substrate compensation is established according to the specimen and the reference weights and sizes [55]. The latter has got a profound importance in precise magnetometry, since due to the inductive nature of the coupling in SQUID magnetometers the strength of the coupling (i.e. the magnitude of the moment reported by the magnetometer) depends not only on the mass of the specimen but also on its vertical (along the sample holder) and radial (perpendicular to the sample holder) extent. In this study we discuss only the results obtained for the in-plane ($H \perp c$) samples orientation, which is the easy direction of $M$ for the samples in question [7,11,18]. The second challenge is connected with an inherent difficulty of a straightforward determination of $x_M$ from low-$T$ $M(H)$, which even at the lowest $T = 2$ K and $H > 50$ kOe continues to increase (slowly) with $H$. This issue is addressed further in the text.

We start from Fig. 11 presenting an exemplary $M(H)$ datasets measured at room temperature (RT) for samples S660 (low $x$), S605 and S590 (both high $x$). The last one is the layer with the most morphologically developed surface (as indicated by the AFM, Fig. 1e). The results confirm that at RT in each case $M(H)$ assumes a linear character without exhibiting any remnant moment exceeding the average experimental noise level of about $10^{-7}$ emu. It should be noted that such an uncertainty constitutes less than 0.25 and 0.05 % of the high field values of $M$ attained at 2 K for sample S660 and either of the other two samples, respectively. This finding, in combination with the results of the detailed structural analysis presented before, allows us to definitely rule out a presence of any statistically significant contribution from secondary magnetic phases, pointing to random Mn distribution in GaN as the sole source of the samples' magnetization.

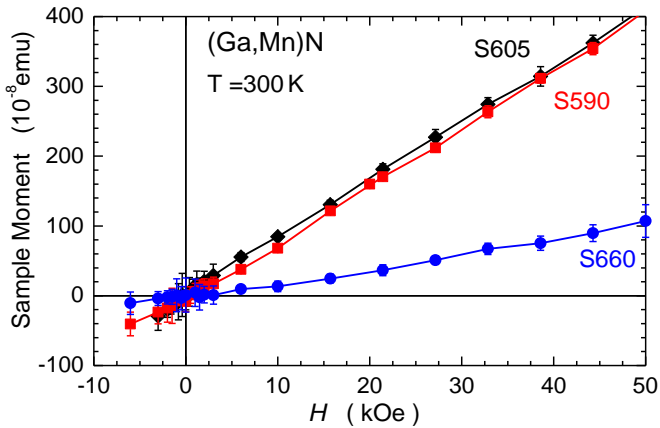

**Fig. 11.** (Color online) Magnetic field ($H$) dependence of the magnetic moment at room temperature for (Ga,Mn)N layers S660 (circles), S605 (diamonds), and S590 (squares). All the presented specimens have nearly the same area of 0.18 cm$^2$. Sizable magnetic contribution specific to the substrate has been subtracted. The experimental uncertainty of the data points is given as the standard deviation values reported by the SQUID magnetometer.

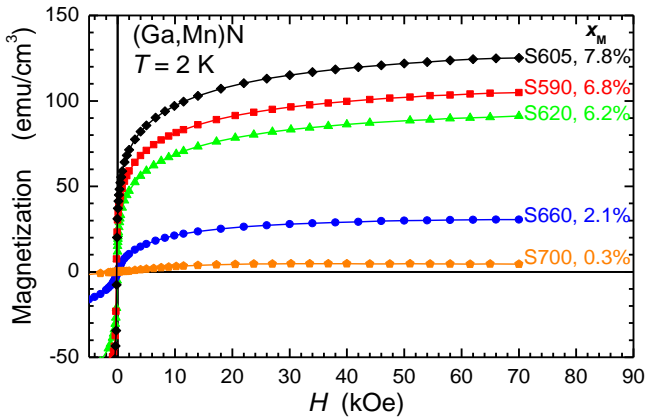

**Fig. 12.** (Color online) Influence of the growth temperature ($T_g$) on the low temperature ($T = 2$ K) magnetic properties of (Ga,Mn)N. All the presented magnetization curves are obtained for the in-plane configuration, i.e. with the magnetic field $H$ aligned perpendicular to the wurtzite $c$ axis. Established according to the method visualized in Fig. 13 magnitudes of Mn concentration in these samples ($x_M$) are also given.

The dependence of the magnitude of the low temperature ($T = 2$ K) $M(H)$ on $T_g$ for all the layers from this study is outlined in Fig. 12. Here, the most representative, the central specimens of each sample are shown. A massive, more than twentyfold increase in $M$ is seen as the $T_g$ is lowered from 700 to 605 °C, confirming the very strong influence of the $T_g$ on the Mn incorporation into the GaN matrix, and so on the magnetic properties of the resulting material. Interestingly, a further reduction of $T_g$ to 590 °C results in a reduction of $M$ indicating that for the selected in this study magnitudes of BEPs of all three elements $T_g = 605$ °C is the optimum growth temperature to obtain a high quality high-$x$ (Ga,Mn)N. Furthermore, what needs to be underlined here, the not optimal growth conditions at $T_g = 590$ °C, which as evidenced by AFM and TEM, result in a considerable surface warping, except of somewhat reduced $x$, have no detectable detrimental influence on the magnetic constitution of this layer. The clearly visible tendency to saturation for $H \rightarrow 70$ kOe allows us to establish $x_M$ for these layers. The employed method is outlined below and the corresponding magnitudes of $x_M$ are given in Fig. 12 and in Table I.

The temperature evolution of $M(H)$ in this series of (Ga,Mn)N layers, taking two samples: S660 (low $x$) and S605 (high $x$) as the examples, is depicted in Fig. 13. Interestingly, at these scales both samples look similar: their linear $M(H)$ at RT remains so down to about 50 K. A sizable curvature appears only below some 20 K and a very pronounced s-like shape of $M(H)$ with a tendency to saturation at the lowest temperatures is exerted. However, it has to be mentioned that $M$ does not really saturate, even at strongest applied fields of about 70 kOe $M(H)$ retains a convex character - it does not really flatten completely. Nevertheless, the high field part of $M(T = 2$ K$)$ permits to assess $x_M$ in these samples with a reasonably high precision.

To this end we model the experimental $M(H)$ within the frame of the group theoretical model for *noninteracting* $Mn^{3+}$ cation including explicitly contributions from the crystal field of the tetrahedral $T_d$ symmetry, the trigonal distortion along the GaN hexagonal $c$ axis, the static Jahn-Teller distortion of the tetragonal symmetry, the spin–orbit interaction, and the Zeeman term [11,18,57,58]. In such an approach $x_M$ is the scaling factor used to align the modelled $M(H)$ (the dashed lines in Fig. 13) with the experimental points in sufficiently wide range of $H$. It is important to note that the perfect match is obtained at $H > 30$ kOe for both dilute (paramagnetic at 2 K) and concentrated (ferromagnetic up to about 12 K). Obtained in this way values of $x_M$ are given in Fig. 12. Although the detailed dependence of $x$ on $T_g$ will be discussed later, now we note that the obtained here a very good agreement with other estimations of the Mn content points, in agreement with XANES, to a relatively small concentration of $Mn^{2+}$ ions for which antiferromagnetic interactions result in $x_M < x$ [19].

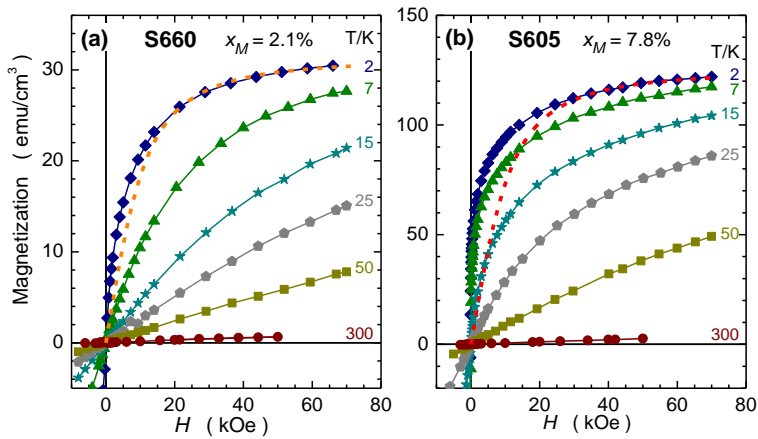

**Fig. 13.** (Color online) (solid symbols) Isothermal in-plane $M(H)$ curves for (Ga,Mn)N samples S660, panel **(a)** and S605, panel **(b)**. The dashed lines represent model calculation within the frame of the group theoretical model for noninteracting $Mn^{3+}$ cations with Mn (magnetic) concentration ($x_M$) as the only free (multiplicative) parameter.

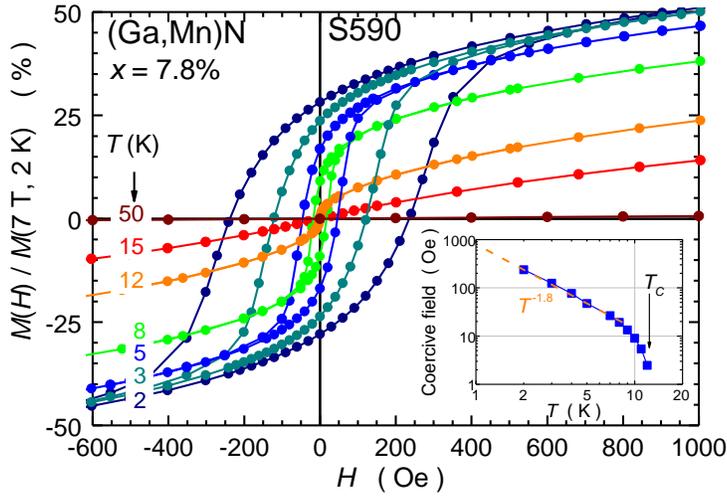

**Fig. 14.** (Color online) A weak field part of isothermal magnetization curves $M(H)$ at selected temperatures for sample S590. The chosen relative representation for $M(H)$ indicates that at 2 K the sample assumes 50% of its 70 kOe saturation value already at $H = 1$ kOe. The inset shows temperature dependence of the magnitude of the coercive field.

Noteworthy, the theoretical modeling gives also a reasonable account for the weak field $M(H)$ for the low-$x$ sample, while a sizable difference is seen for the high-$x$ sample. The vastly enlarged magnitude of the experimentally established $M(H)$ over the theoretical line is indicative of a presence of FM ordering for such high $x$. This FM behavior is further evidenced in Fig. 14 where a weak field part of the low-$T$ $M(H)$ for sample S590 is presented. We add that similarly to the results presented in ref. 7 the width of the nonreversible part of $M(H)$s as well as their squareness increase on lowering $T$. The inset to Fig. 14 documents a rapid, nearly as fast as $1/T^2$ increase of the magnitude of $H_C$ on lowering $T$ in this sample. Importantly, it has to be noted that open hysteresis curves are observed only at the lowest temperatures. Above a specific temperature for each sample, which we identify with the Curie temperature, despite exhibiting still sizable nonlinearities, none of the investigated samples shows any history dependent effects.

A magnitude of $T_C$ is probably the most important figure of merit of DFS, determining in particular their possible technological relevance. It has to be noted that despite its significance no definitive method of its precise determination have been established so far. For example, the use of time honored methods of standard or modified Arrott plots [59] requires *a priori* a knowledge of the critical exponents (a universality class) specific to the investigated material. This condition is sizably relaxed in the Kouvel-Fisher method [60], shown to work remarkably well in (Ga,Mn)As [61]. However, this method does not work in dilute systems with short range magnetic interactions, like (Ga,Mn)N. Particularly in such materials, following the Griffiths suggestion [62], the presence of preformed ferromagnetic clusters may sizably smear the phase transition and shift up the temperature at which the first signatures of FM are detectable. Additional smearing may also result from macroscopic inhomogeneities of spins distribution [7] inherent to virtually all real magnetic alloys, complicating the matter even further. Now, since in all homogeneous magnetic alloys $T_C$ is a function of the concentration of the magnetic element, both microscopic and macroscopic variations of $T_C$ are expected in our samples, necessitating a more cautious approach to its determination.

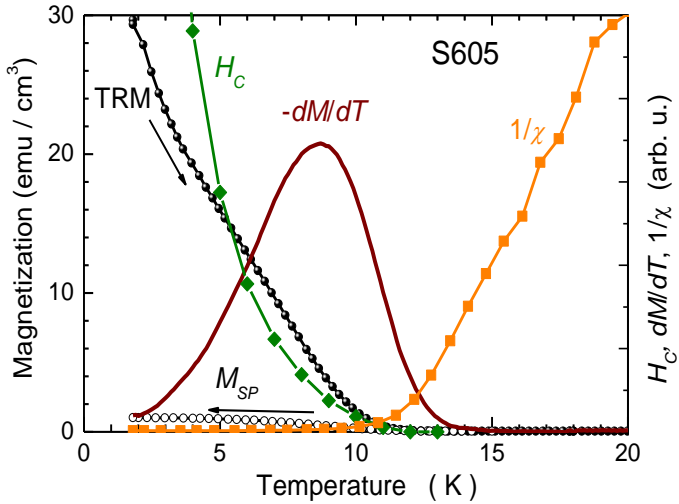

**Fig. 15.** (Color online) Critical behavior of layer S605 and Curie point determination from the inverse magnetic susceptibility ($1/\chi$), coercive field ($H_C$), thermoremanent (TRM), and spontaneous magnetization ($M_{SP}$), as well as from the position of the inflection point on temperature dependence of magnetization $M(T)$ (indicated here as the maximum of numerically established $-dM/dT$ measured at $H = 1$ Oe).

There are a few practical methods of $T_C$ assessment which are frequently in use. Five of them are exemplified in Fig. 15 for sample S605. We see that similarly to findings presented in Fig. 4 of ref. 7 there exists a very good agreement between the value of $T_C = 13.0 \pm 0.3$ K determined from an extrapolation of the coercive field ($H_C$), inverse of the magnetic susceptibility ($1/\chi$), thermoremanence (TRM), and the spontaneous moment ($M_{SP}$) magnitudes towards zero. For the TRM the specimen is cooled down in a field of an order of 1 kOe and at the base temperature ($T = 1.8$ K) the field is quenched (using the soft quench of the superconducting magnet). Under these conditions the TRM is collected on increasing temperature until the signal drops to zero (that is decidedly below the noise level). The $M_{SP}$, perhaps the least commonly performed measurement, is commenced right after the TRM one during cooling down the sample at the same $H \simeq 0$ conditions. This measurement yields the direct information on both the temperature at which the long-range-coupled spontaneous moment forms and on the magnitude of this moment [63-65]. This method is particularly effective in detection of even a very weak spontaneous moment in magnetically complex materials since all the other magnetic constituencies yield no magnetic contribution when are cooled down from the above of their characteristic temperatures at $H = 0$. This remark is particularly relevant to materials containing a sizable portion of spins coupled in superparamagnetic clusters exhibiting blocking properties [63,66,67]. More importantly the $M_{SP}$ results play a decisive role in the determination which part of the decaying TRM at the critical region should be taken as the indication of the transition point, or how to extrapolate it to zero. This technical issue gets particularly important when in contrast with the concave Brillouin behavior, expected in a ferromagnet according to the Weiss mean field theory, the convex shape of the slowly decaying TRM - expected for a sizably dilute magnetic material in which FM order is formed in percolation fashion [68] - renders such determination greatly unreliable.

The convergence of TRM, $M_{SP}$, $1/\chi$, and $H_C$ methods presented in Fig. 15 is the common feature observed in all our (Ga,Mn)N samples in which $x \gtrsim 5\%$ is sufficiently high to allow observation of the signatures of magnetic coupling in the available temperature range, $T > 2$ K. It is so because all these methods are sensitive to the upper bound of the $T_C$ distribution within the investigated specimen. Therefore the magnitude of $T_C$ established by either of these approaches corresponds to the highest statistically significant $T_C$ in the specimen ($T_C^{max}$), that is to a temperature at which first statistically relevant long-range coupled regions are formed. On the other hand, the fifth method, which locates the Curie transition at this temperature at which $M(T)$ dependency has an inflection point (IP method) yields a sizably different magnitude of $T_C$. This is because this method is sensitive to the statistically most significant mean $T_C$ value in the specimen ($T_C^{mean}$).

## 3. Discussion

The first main result of this study – the sizable dependence of Mn incorporation into GaN on the growth temperature – is depicted in Fig. 16. All SIMS, XRD and SQUID data sets agree in their indications that the reduction of $T_g$ from 700 $^o$C to about 600 $^o$C increased Mn content at least ten-fold up to about 10%, changing in a rather smooth manner. Undoubtedly, an enhanced Mn incorporation above 10% could be realized by further fine tuning of both the growth and the N/metals flux ratios at around 600 $^o$C or little above. Importantly, our XRD, HRXRD, and particularly SQUID studies preclude the presence of any alien crystallographic phases or truly large variations of $x$ in all of these layers, strongly indicating that single phase random DFS magnetic alloys can be grown even outside of the commonly accepted envelope of the growth parameters for a given parent semiconductor. This fact has got a profound implication for possible functionalization of DFS, as our studies show that a growth of complex multilayers structures necessitating the presence of magnetic layer of different micromagnetic properties, say for spin-valve or resonant tunneling devices, can be largely simplified, since the required diversity can be accomplished by changing mainly the $T_g$. Moreover, this dependency is so strong that even a much finer variations of $T_g$ than the major coarse steps between these five growths are clearly seen in Fig. 16. The data presented in this figure are obtained for various pieces cleft from differen zones of substrates and they are plotted against their real $T_g$, which has been established individually for each specimen by temperature mapping across the substrate, as exemplified in Fig. 10. As indicated there, the variation of $T_g$ can be as high as 12 $^o$C between the center and the edge of the substrate and the results plotted against the real $T_g$ correlate very nicely with the general trend established by the main changes of $T_g$. It is also important to notice an exceptionally close match between the magnitudes of $x$ established by the three methods for the material grown above 600 $^o$C. This is yet another indication that Mn in these layers assumes predominantly +3 oxidation state since the antiferromagnetic coupling specific to $Mn^{2+}$ state would diminish $x_M$ ($x$ from magnetization) with respect to SIMS and XRD findings. Perhaps this is the case of the material grown below 600 $^o$C, since the $x_M$ of these specimens is visibly reduced with respect to the concentrations established by SIMS and XRD.

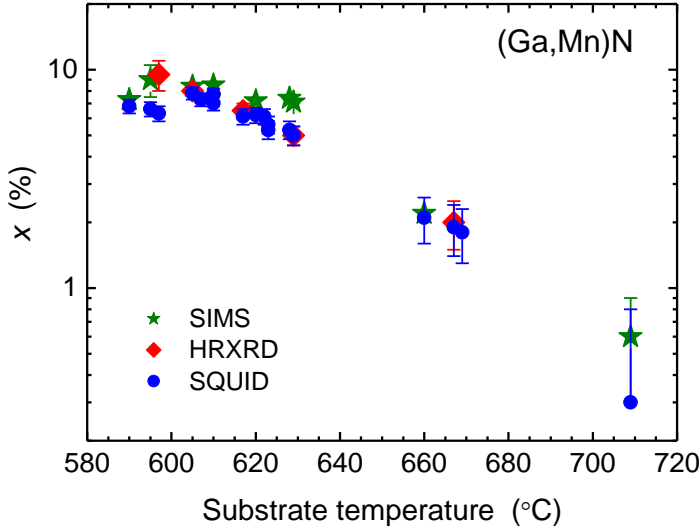

**Fig. 16.** (Color online) Mn content ($x$) determined by secondary ion mass spectrometry (SIMS), high resolution X-ray diffraction (HRXRD), and magnetometric studies as a function of the substrate temperature established individually for each specimen depicted in this figure.

Turning back to the importance of the $T_g$ on the determination of the micromagnetic properties of (Ga,Mn)N we concentrate now on its influence on $T_C$ and to this end only the three layers grown at the lowest $T_g$ are investigated, since only for these layers $x > 5\%$, and so $T_C > 2$ K. According to the already established vertical and in-plane macroscopic inhomogeneities of spin distribution we expect sizable differences between $T_C^{mean}$ and $T_C^{max}$ ($\Delta T_C$) in our layers. We actually move a step further and suggest that the magnitude of $\Delta T_C \equiv T_C^{max} - T_C^{mean}$ can be taken as an additional practical measure of the microscopic magnetic inhomogeneity of the material.

First, we illustrate, using only partially processed experimental data that indeed sizable variations across the substrate are found in our samples. Numerically calculated first derivatives of the $M(T)$ dependencies obtained for various specimens cleft from layer S620 are presented in Fig. 17. From such a derivative one gets readily the position of the inflection point on the $M(T)$ dependency, the $T_C^{mean}$, that is the temperature at which a minimum on $dM/dT$ is seen. The labels in this figure correspond to the coding of the specimens which were cleft from this layer and the whole sample layout is presented in Fig. 10. It is immediately seen that the established values of $T_C^{mean}$ span a sizable range running from 4.1 to 6.4 K, or its variation exceeds 40% of the mean value. Remarkably, by a comparison with Fig. 10 we find that the lowest magnitude of $T_C^{mean}$ is seen for the specimen obtained right from the rim of the substrate with the lowest $T_g$ in this set (specimen "D2", $T_C^{mean} = 4.1$ K for $T_g = 630$ °C), whereas the largest $T_C^{mean}$ is detected for specimen from the very center of the substrate (specimen "C", $T_C^{mean} = 6.3$ K and $T_g = 620$ °C – the intended $T_g$ for this layer). This is actually a very strong effect, in particular that obtained on a material for which all typical measures applicable for MBE growth of nitrides have been undertaken to assure the most efficient heat transfer and temperature uniformity.

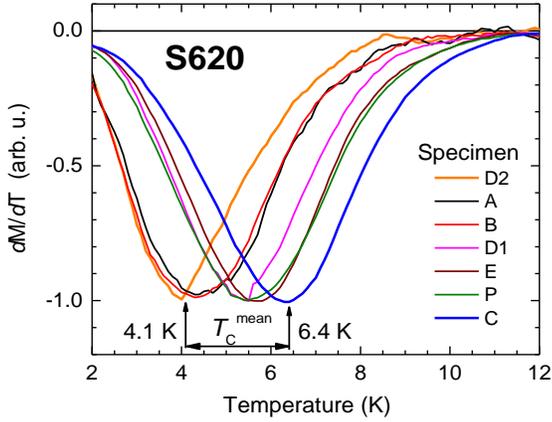

**Fig. 17.** (Color online) Normalized numerically calculated first derivatives of the temperature dependent magnetization ($dM/dT$) for several specimens originating from layer S620. The coding of the specimens corresponds to the layout of this layer presented in Fig. 10. The arrows indicate the lowest (4.1 K) and the largest (6.4 K) mean Curie temperature ($T_C^{mean}$) in this ensemble

The results presented above are by no means specific to this layer. Neither they are characteristic for the IP method of the $T_C$ determination. Magnitudes of the Curie temperatures established by both IP and TRM methods ($T_C^{mean}$ and $T_C^{max}$, respectively) for all the specimens researched in this study of sufficiently high $x$, that is originating from layers S620, S605 and S590 are collected in Fig. 18. Table I lists both magnitudes of $T_C$ established in the central zone of each layer. First of all, we notice that the data collected in this figure follow the same trend already exhibited for $x(T_g)$ dependency (Fig. 16). $T_C$ is increasing with lowering $T_g$ down to about 600 °C and starts to fall off below. Furthermore, we note that this dependency can be traced with nearly a single-degree resolution and that this general relationship is perfectly followed from layer to layer depending only on the actual $T_g$ that a particular part of the substrate is experiencing during the growth. More importantly, an observation of such a clearly defined dependency of $x$ and $T_C$ on local (actual) $T_g$ unambiguously indicates that these are the inhomogeneities of the $T_g$, not of the molecular fluxes, that are predominantly responsible for the properties of the material.

The observed peak values of $x$ and $T_C$ at around 605 °C come rather as a surprise. According to numerous growth reports discussed in the Introduction at such a reduced $T_g$, as for GaN, one expects an increased incorporation of oxygen and a generation of numerous structural and point defects - all reducing the Mn oxidation state to +2 (this one couples antiferromagnetically) and/or stimulation of other magnetic phases nucleation. Our extensive characterization rules out these effects at least up to a level that would have a detrimental influence on (ferro-)magnetism of these layers. In fact, in absolute numbers, Mn concentration amounts to $4 \times 10^{21}$ cm$^{-3}$, so having a joint concentration of donor-like centers exceeding even $10^{20}$ cm$^{-3}$ may actually come unnoticed in (Ga,Mn)N. However, a further reduction of $T_g$ to 590 °C actually causes a small reduction of $x$ and $T_C$. At present it remains unknown whether this reduction is correlated with the development of the surface warping, or with an enhanced Mn outdiffusion - suggested by evidenced by SIMS sizable reduction of $x$ towards the surface in this layer. Nevertheless, our findings indicate that perhaps other previously unexplored areas of the growth parameter space could be successfully tested, leading to a further progress in Mn incorporation in nitrides.

Now we take advantage that the two different methods for $T_C$ determination are used. We notice that both methods yield a very similar $T_C(T_g)$ dependency. We strongly underline this general consistency. However, there is a finer structure embedded in these relationships, best seen when the difference between $T_C^{max}$ and $T_C^{mean}$ is plotted. As suggested before, the magnitude of $\Delta T_C$ informs us on a degree of the $T_C$ distribution within the specimen. We find this degree rather large in our samples. On average $\Delta T_C \simeq 4$ K and is much larger than it was observed for similar MBE samples grown about 4 times faster (200 nm/h) and at higher $T_g$ [7]. Then, values below 2 K were recorded. We assign this enhanced distribution of $T_C$ to the additional sizable vertical gradient of $x$ established in these layers by SIMS profiling (Fig. 2). Our conjecture is supported by the fact that the lowest magnitude of $\Delta T_C$ is obtained for the central piece of S605 layer, for which depth dependence of $x$ is found far weaker than in the other two high-$x$ layers.

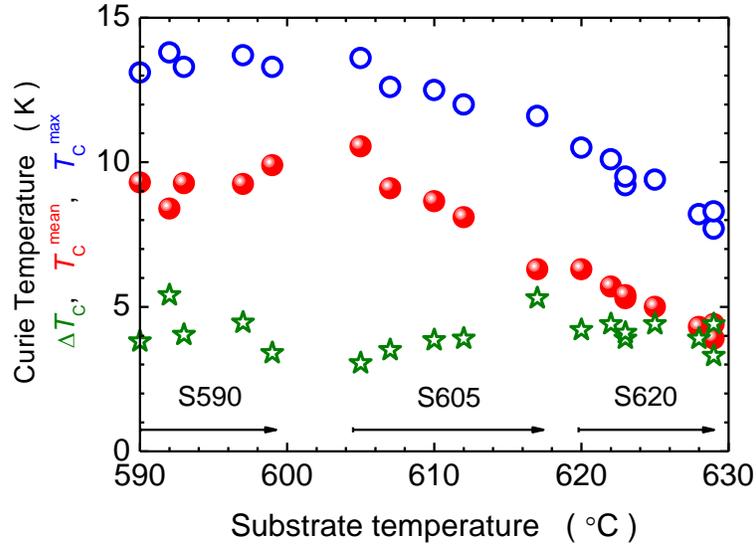

**Fig. 18.** (Color online) Curie temperature ($T_C$) as the function of the substrate temperature assigned individually for each specimen according to substrate temperature mapping as in Fig. 10. The magnitudes of $T_C$ are determined from the extrapolation of thermoremanence (TRM) to zero ($T_C^{max}$, open symbols) and from the position of the inflection point (IP) of the temperature dependent magnetization measured at 1 Oe ($T_C^{mean}$, solid symbols). Stars represent the difference $\Delta T_C = T_C^{max} - T_C^{mean}$.

Finally, we note that Figs. 16, 17, and 18 illustrate very clearly an importance of one of the main prerequisites of characterization measurements that the actual measurements are to be performed either on closely matching material (even on physically the same specimens, if that is possible) or a relevant parameter responsible for measurable variation of the physical properties should be directly taken into account. This fact seems to be the most plausible explanation of the sizable discrepancies in the Mn concentration reported previously by some of the present authors (cf. Fig. 4 in ref. 5), as all the methods employed there (SIMS, XRD, and magnetometry) probed physically different pieces, originating from different zones of the parent sample.

## 4. Conclusions

In this paper, we have investigated a range of Ga$_{1-x}$Mn$_x$N layers grown by MBE with manganese concentration $x \leq 10\%$ set predominantly by changing the growth temperature between 590 and 700 °C. Despite of an obvious deviation from the stoichiometric growth point with the change of $T_g$, high quality layers have been obtained as determined by a set of experimental methods, including AFM, SIMS, HRTEM, and XRD. We relate this to the lateral-growth-promoting role of Mn which acts as surfactant. These measurements reveal the, absence of crystallographic phase separation, clusters or nanocrystals, and indicate an increase of surface roughening with lowering $T_g$. The characterization effort has been combined with an extensive in-depth magnetic studies. In particular, our SQUID, as well as XANES with DTF calculations, results point to the predominantly +3 configuration of Mn in GaN and accordingly the ferromagnetic phase has been observed for layers with $x > 5\%$ at $3 < T < 10$ K, in agreement with the already elaborated trends in (Ga,Mn)N layers grown much closer to the typical growth conditions for GaN.

Within such an approach $x$ can be changed more than tenfold, and the increase of $x$ occurs solely on the account of the reduction of $T_g$. Like in the case of In, the vapour pressure of Mn is much higher than that of Ga and its desorption from the surface is reduced during the growth performed at relatively low temperature. A detrimental side of a substantial lowering of $T_g$ is the reduction of surface mobility of the atoms what seems to be the main reason of the surface warping at $T_g < 600$ °C. However, we do not detect other obvious imperfections as stacking faults and additional macrodefects, except of rare nm-sized voids. This in turn seems to be a positive aspect of very low growth rates exercised by us in this study as the adatoms are given an extra time to assume positions commensurate with the lattice. It is worth noting that the overall roughness of low $T_g$ surfaces stays below 4 nm what does not disqualify this material from further functionalization.

Along with the increase of $x$ the magnitude of $T_C$ follows, and it has been shown that these two quantities correlate with the separately established local temperature of the material within a nearly a single degree resolution. Most likely the main reason of the spatial variation of $T_g$ in our samples is a nonuniform heat transfer from the heater to the substrate. The effect can be so substantial that the local variation of $T_C$ across about 1 inch substrate may amount up to 40%. Alternatively, a strong relation of $T_C$ on $x$ in (Ga,Mn)N, and so on $T_g$, turns magnetic studies into a very efficient tool for precise determination of the temperature and its distribution across the substrate, an indispensable piece of information required for correct settings and elaboration of the whole growth protocol in such a nonequilibrium growth technique as MBE is, particularly for ternary compounds in which metal species differ in almost every aspect of their growth related parameters determining the kinetics of the growth. Apart from practical aspects, the accumulated in the study evidences strongly necessitate a further search for methods which would eliminate thermal inhomogeneities of the substrate during the growth. We also show that the determination of $T_C$ by two different methods, each sensitive to different moments of $T_C$ distribution within the specimen, may serve as a tool for quantification of spin homogeneity within the material.

**Acknowledgements**

This work has been supported by the National Science Centre (Poland) through grants FUGA (2014/12/S/ST3/00549) and OPUS (2013/09/B/ST3/04175). We gratefully acknowledge J. Sadowski for valuable comments and discussion, SOLEIL for provision of synchrotron radiation facilities, and dr. Emiliano Fonda for assistance in using beamline SAMBA.

**Declarations of interest**: none